# Orchestr.AI: A Unified Workflow for Training and Benchmarking of Machine-Learned Interatomic Potentials

*Zain Ul Abideen[1,3], Mario Fernández-Pendás[2], Muhammad Usman[1], Camilo Rodríguez Quintero[1], Masuma Suleymanova[1], Ignacio Arganda-Carreras[3,4,5,6], Ivan Infante[1,6]**

[1] BCMaterials, Basque Center for Materials, Applications, and Nanostructures, UPV/EHU Science Park, Leioa, 48940 Spain

[2] Departamento de Métodos Cuantitativos, Facultad de Economía y Empresa, University of the Basque Country (UPV/EHU), Bilbao, Spain

[3] Department of Computer Science and Artificial Intelligence, Faculty of Informatics, University of the Basque Country (UPV/EHU), Donostia-San Sebastián, Spain

[4] Donostia International Physics Center (DIPC), San Sebastián, Spain

[5] Biofisika Institute (CSIC-UPV/EHU), Leioa, Spain

[6] Ikerbasque Basque Foundation for Science, Plaza Euskadi 5, Bilbao, 48009 Spain

E-mail: ivan.infante@bcmaterials.net

**Funding:** UNICORN (Grant Agreement No. 101098649) and the Marie Skłodowska-Curie Actions Doctoral Network Track The Twin (Grant Agreement No. 101168820)

**Abstract.**

Machine-learning force fields enable atomistic simulations at near-reference accuracy and substantially reduced computational cost, but their development remains fragmented across software frameworks with incompatible data formats, configuration schemes, dependencies, and execution procedures. Orchestr.AI addresses this limitation through a unified workflow that coordinates different MLIP software frameworks, referred to as engines, for data preparation, model training, fine-tuning, evaluation, and deployment. A common YAML interface defines each experiment, while engine-specific environment profiles preserve compatible software stacks and avoid dependency conflicts. The platform currently supports five MLIP model architectures, namely SchNet, PaiNN, NequIP, Allegro, and MACE, through their corresponding software implementations, providing automated configuration mapping and standardized dataset handling across these frameworks. Benchmarks on the rMD17 and 3BPA datasets show that Orchestr.AI reproduces the expected performance trends while enabling controlled cross-engine comparisons. Repeated executions with fixed configurations and engine profiles produce highly consistent results, demonstrating practical workflow reproducibility. Beyond training, Orchestr.AI provides modular architectural interfaces that connect trained models to ASE-based simulations, batch inference, epistemic uncertainty estimation, calibration, and active-learning workflows. The platform therefore provides a reproducible and extensible infrastructure for developing, comparing, and deploying machine-learning interatomic potentials, which we validate here through cross-engine accuracy and repeatability benchmarks.

## 1. Introduction

Atomistic simulation has long been governed by a compromise between physical accuracy and computational reach. First-principles electronic-structure calculations, ranging from density functional theory (DFT) to correlated wave-function approaches, can describe chemically diverse environments without relying on a predefined functional form, but their computational cost restricts the accessible size and time scales. Empirical force fields enable simulations of much larger systems and longer trajectories, yet their fixed parameterizations can become unreliable when the relevant chemical environments extend beyond those anticipated during model construction. Machine-learned interatomic potentials (MLIPs), also commonly referred to as machine learning force fields (MLFFs), have reshaped this compromise by learning potential-energy surfaces directly from electronic-structure data. Once trained, these models can approach the accuracy of their reference method at a fraction of its computational cost, opening atomistic simulation to systems and processes that would otherwise remain inaccessible [1–5].

The development of MLIPs has progressed rapidly from models based primarily on invariant local representations to architectures that encode increasingly rich geometric information. Continuous-filter and message-passing models such as SchNet established the effectiveness of neural representations learned directly from atomic coordinates [6–8]. Other invariant or geometry-aware approaches, including DimeNet, GemNet, and AIMNet, further enriched atomic representations through angular, higher-order, or learned environment descriptors [9–11]. Direction-aware architectures such as PaiNN[12] subsequently incorporated rotationally equivariant features, while NequIP and Allegro[13,14] demonstrated that enforcing Euclidean symmetries throughout the network can substantially improve data efficiency and force accuracy. MACE further combines equivariance with systematically constructed higher-body-order representations, providing a flexible balance between accuracy and computational efficiency[15]. Equivariant architectures such as TensorNet, eSEN, and SevenNet, together with equivariant transformer models including EquiformerV2 and SO3krates have further expanded the range of geometric representations and interaction mechanisms available for atomistic learning.[16–20] More recently, pretrained and foundation interatomic potentials, including M3GNet, CHGNet, MACE-MP-0, PET-MAD, GRACE, have extended this trajectory toward models spanning broad regions of chemical and materials space, with the possibility of using them directly in specific applications or improving them through fine-tuning.[21–25] As a result, researchers can now choose among a growing number of capable architectures with different trade-offs in accuracy, training cost, inference speed, scalability, and transferability.[1,6,7,12,14,15,26]

Reliable MLIPs depend on data curation, units, partitions, model and training settings, and deployment as well as architecture. Choices at any stage can change apparent performance: a nominally identical parameter, such as a cutoff radius, loss weight, or learning-rate schedule, may also be represented or interpreted differently by different software packages. Consequently, comparisons that appear equivalent at the level of a published table may conceal substantial differences in data processing, configuration semantics, or execution.

These difficulties are amplified by the fragmentation of the MLIP software ecosystem. SchNet and PaiNN are commonly accessed through SchNetPack, whereas NequIP, Allegro, and MACE are distributed through distinct codebases with their own configuration schemas, data formats, command-line interfaces, dependency requirements, and distributed-training procedures.[6–8,12–15] Other widely used MLIP codes and frameworks, including DeePMD-kit, TorchMD-Net, MatGL, and SevenNet, similarly provide their own data-processing, configuration, training, and deployment workflows.[18,27–29] Across these different software ecosystems, rapidly evolving dependencies, including PyTorch, CUDA, and e3nn, can change model definitions, supported options, and default behaviors between releases.[30] Changing dependencies and configuration semantics make earlier calculations difficult to reconstruct and cross-engine comparisons difficult to control.[31–33]

Several tools simplify individual stages of MLIP development. The Atomic Simulation Environment (ASE) provides a common representation of atomic structures and a calculator interface for atomistic simulations, while leaving model training to external packages[34]. KLIFF supports potential fitting and uncertainty analysis within its own framework and through OpenKIM interfaces[35,36] , while SPARC automates active learning for reactive systems [37], using DeePMD-kit for potential training alongside ASE and PLUMED[38,39]. Their main contributions lie in simulation, fitting, or iterative data generation rather than in coordinating training across independently maintained MLIP engines.

Other platforms unify models or their evaluation, but at a different layer. The JAX-based mlip library and apax implement model architectures within their respective frameworks, while AI2Pot brings several potential types into a shared PyTorch-compatible framework with optimized operators. MLIP Studio provides a common interface for applying and comparing pretrained potentials, and MLIPAudit standardizes their downstream evaluation. These approaches therefore bring models into a shared implementation or bring trained models together for use and assessment [40–46] . Workflow-level orchestrators such as IPSuite, psiflow, wfl, and autoplex also coordinate multi-engine training, subsampling, and active-learning tasks through programmatic Python APIs or data-versioning pipelines [47–50].

Against this background, our Orchestr.AI introduces a complementary, configuration-centric form of interoperability. Its defining feature is that unification occurs at the level of the scientific experiment through a declarative master YAML translation layer and fixed environment profiles rather than within a newly implemented machine-learning backend or custom Python workflow scripts. SchNet and PaiNN continue to be trained through SchNetPack, while NequIP, Allegro, and MACE retain their native implementations, dependencies, configuration options, and update cycles. Above these engines, Orchestr.AI provides a common master YAML specification that is translated through explicit and validated mappings into the configuration required by each backend. It couples this engine-agnostic experiment description with isolated software environments, standardized dataset validation and conversion, controlled train-validation-test partitions, automated single- and multi-GPU execution, and consistent organization of checkpoints, predictions, logs, and evaluation metrics. Engine-specific overrides remain available when a model requires parameters that cannot be represented by the shared schema, preserving expert control without requiring separate workflows for routine use. The present work also evaluates the core training, benchmarking, and reproducibility capabilities of Orchestr.AI. The rMD17 benchmark is used to examine cross-engine performance and repeated-run stability, while temperature-dependent and conformational tests on 3BPA probe generalization beyond the training distribution. The objective is not to claim state-of-the-art accuracy for the individual architectures, nor to reproduce numerical results obtained with historical software versions exactly. Instead, the study asks whether heterogeneous, actively maintained MLIP engines can be configured, executed, and evaluated through a transparent and reproducible workflow while retaining their expected performance characteristics, leaving empirical case studies of the data-consolidation, fine-tuning, and active-learning modules to dedicated follow-up studies.

## 2. Methods

**Figure 1** presents the architecture and end-to-end workflow of Orchestr.AI. Starting from reference data and a master YAML experiment specification, the platform validates and converts the input data, identifies the requested operating mode, and passes the experiment through an engine adapter layer. This layer handles templates, parameter mappings, compatibility checks, and optional engine-specific overrides before generating the native configuration required by each MLIP framework. The resulting training, fine-tuning, or benchmarking tasks are dispatched to the appropriate software environments through engine-specific execution wrappers. Configurations, checkpoints, logs, predictions, and evaluation metrics are subsequently collected within a standardized output structure that supports consistent analysis across engines. Because engine-dependent operations are isolated within modular adapters, additional frameworks can be incorporated without modifying the common workflow, while several supported engines can be trained and benchmarked from the same experiment definition. The following subsections describe the principal components of this workflow, from data preparation and training to cross-engine benchmarking, fine-tuning, uncertainty estimation, and active learning.

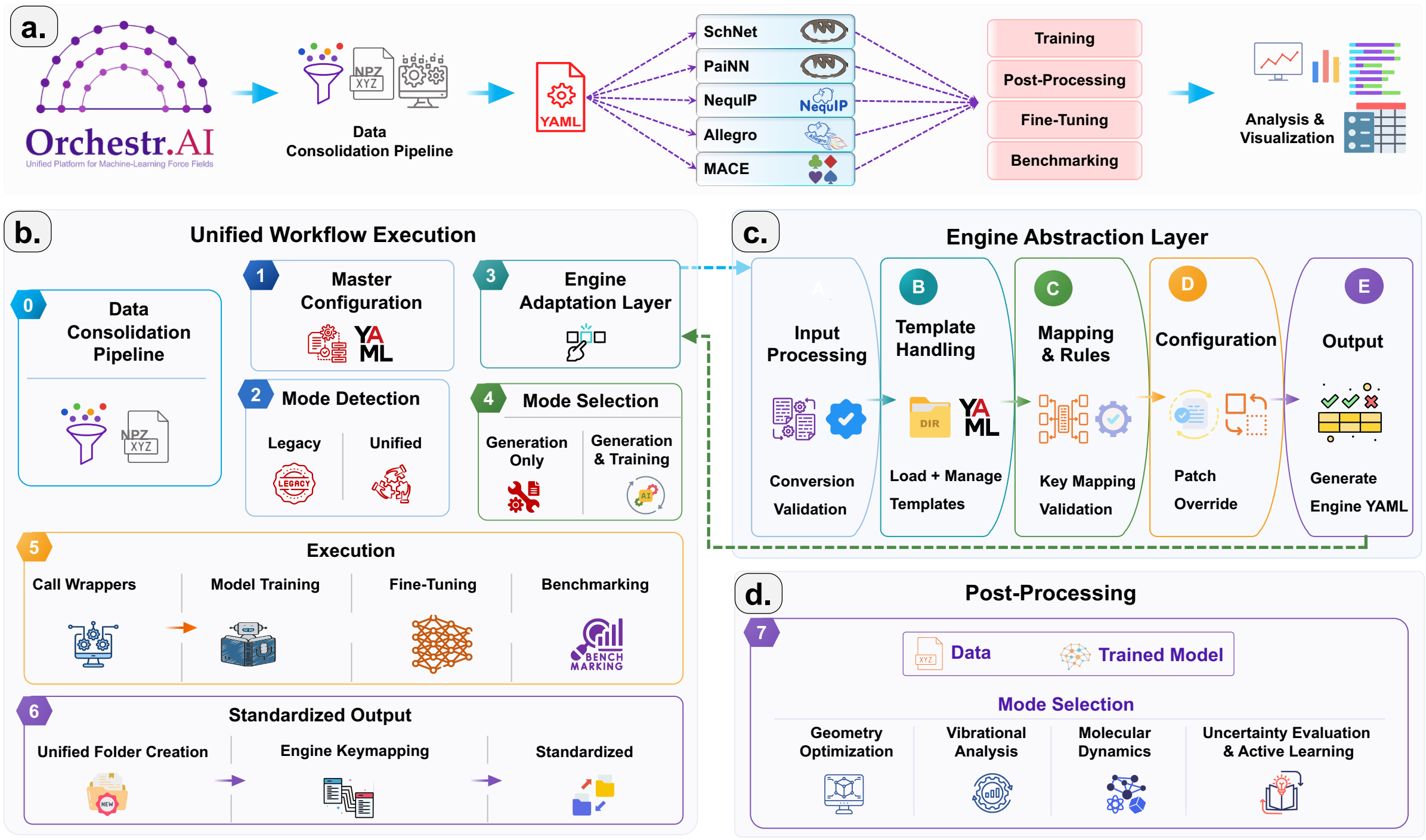

**Figure 1**: **Architecture and end-to-end workflow of Orchestr.AI.** (a) Overview of the platform, illustrating the data consolidation pipeline, unified YAML configuration, supported MLIP model architectures (SchNet, PaiNN, NequIP, Allegro, and MACE), and training, fine-tuning, post-processing, and benchmarking capabilities. (b) Unified workflow execution comprising seven stages (0–6): data consolidation (0), master configuration (1), mode detection (2), engine adaptation (3), mode selection (4), execution (5), and standardized output organization (6). (c) Detailed view of the Engine Abstraction Layer, which implements Stage 3 through five steps: (A) input processing and validation, (B) template handling, (C) parameter mapping and validation, (D) configuration patching and engine-specific overrides, and (E) generation of the native engine YAML configuration. (d) Optional post-processing stage (7), which uses the resulting data and trained models for geometry optimization, vibrational analysis, molecular dynamics, uncertainty evaluation, and active learning.

### *2.1 Data Preparation and Consolidation*

The development of an MLIP begins with reference calculations, but the resulting data are rarely produced in the representation expected by a training framework. Electronic-structure programs and MLIP engines organize atomic structures, energies, forces, units, and metadata differently. Preparing a dataset separately for every engine would therefore introduce repeated conversion steps and create opportunities for nominally identical training datasets to diverge. Orchestr.AI avoids this problem by first transforming the reference calculations into a common intermediate representation, from which all subsequent engine-specific datasets are generated.

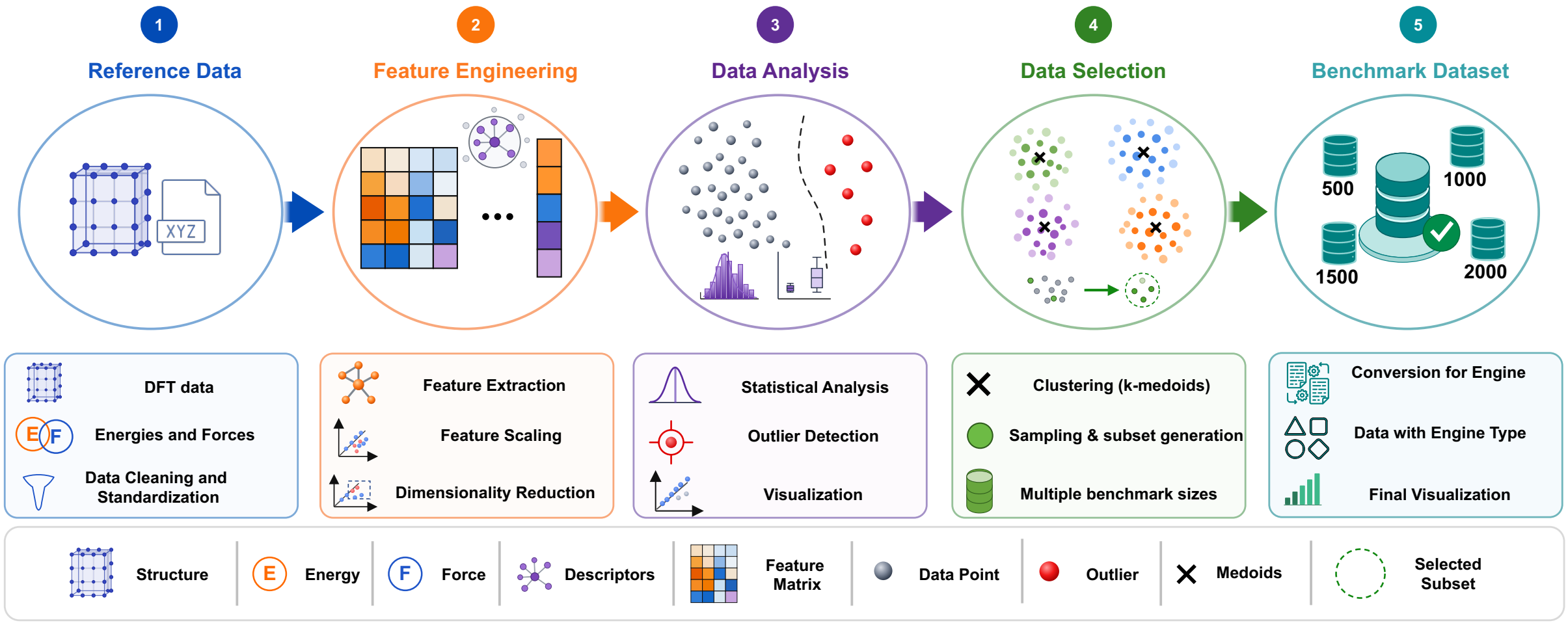

**Figure 2**: **Data-preparation and consolidation workflow implemented in Orchestr.AI.** The workflow consists of five stages: (1) reference data collection, including atomic structures, energies, and forces; (2) feature engineering using reference-property statistics and SOAP descriptors; (3) data analysis, including outlier detection and visualization; (4) data selection through clustering and medoid-based representative configuration selection; and (5) benchmark dataset preparation and export into engine-compatible formats.

The current preprocessing implementation is primarily designed for CP2K data.[51] It can read either a preassembled stacked XYZ dataset containing structures, energies, and forces or separate position and force trajectories. In the latter case, the trajectories are matched and combined, while energies and forces are converted from Hartree and Hartree Bohr$^{-1}$ to eV and eV Å$^{-1}$, respectively. The resulting stacked XYZ file provides a simple internal representation in which each configuration contains the quantities required for MLIP training. Importantly, the remainder of the workflow is not intrinsically tied to CP2K. Outputs from another quantum-chemistry program can be incorporated by implementing a dedicated parser that produces the same intermediate representation, leaving the consolidation and engine-export stages unchanged.

Once the data have been brought into this common form, Orchestr.AI can examine whether they provide a sufficiently consistent and diverse description of the sampled configuration space. As illustrated in **Figure 2**, each configuration is characterized using both reference-property statistics and structural information. Its per-atom energy, mean force magnitude, and force variance describe the energetic and mechanical response of the structure, while a structure-averaged Smooth Overlap of Atomic Positions (SOAP) descriptor generated with DScribe captures the diversity of its local atomic environments [52]. Combining these quantities allows the analysis to distinguish configurations that are unusual either in their geometry or in their associated energy and forces. Because the different components of this representation have unrelated numerical scales and dimensions, the resulting feature matrix is standardized and optionally reduced via principal component analysis (PCA) to balance structural and energetic contributions before further statistical analysis.

The standardized feature space is first examined for atypical configurations using an Isolation Forest with a user-defined contamination level [53]. This unsupervised procedure is useful at this stage because potentially problematic structures such as unconverged electronic-structure steps are not generally associated with predefined labels and may differ from the main dataset in several properties simultaneously. The contamination threshold is user-configurable (or can be disabled) to avoid discarding physically relevant high-energy configurations. The identified configurations can consequently be removed before subset construction, while dimensionality-reduction projections provide a visual representation of the dataset before and after filtering. These diagnostics allow the user to inspect how the preprocessing decisions affect the retained configuration space rather than treating dataset cleaning as an opaque operation.

The filtered dataset can then be reduced to one or more representative subsets of user-defined sizes prior to train–validation–test partitioning. Orchestr.AI applies k-means clustering [54] in the combined feature space and selects, for each cluster, the actual configuration closest to its centroid. This final medoid-selection step is important because a cluster centroid is an abstract point in feature space and does not necessarily correspond to a physically existing atomic structure. Selecting a real configuration from every cluster produces a structurally diverse dataset that can be passed to MLIP workflows. An optional elbow analysis can help estimate a suitable number of clusters, while coverage statistics quantify how closely the selected structures represent the complete dataset. Random subsets can also be generated as baselines, allowing the user to assess whether diversity-based selection provides better coverage than conventional random sampling.

Finally, every selected subset is exported from the same consolidated source into the representation required by the chosen MLIP engine. SchNet and PaiNN receive NPZ files containing atomic numbers, positions, energies, and forces, whereas NequIP, Allegro, and MACE receive extended XYZ files with the required property labels and metadata. In this way, differences in file structure are handled by Orchestr.AI rather than through independent user-written conversion scripts. The data received by the different engines therefore originate from the same configurations and preprocessing decisions, providing a consistent foundation for the cross-engine training and benchmarking described in the following sections. A broader methodological assessment of the consolidation and sampling strategies will be presented separately.

### 2.2 *Unified and Engine Specific Training*

Training the same MLIP experiment with different software frameworks is not simply a matter of changing the model name. Each engine expresses datasets, model parameters, loss functions, hardware resources, and output locations through a different configuration structure. Even parameters that represent the same physical or numerical choice may have different names, locations, or interpretations. Orchestr.AI addresses this problem by separating the scientific definition of an experiment from the native configuration used to execute it.

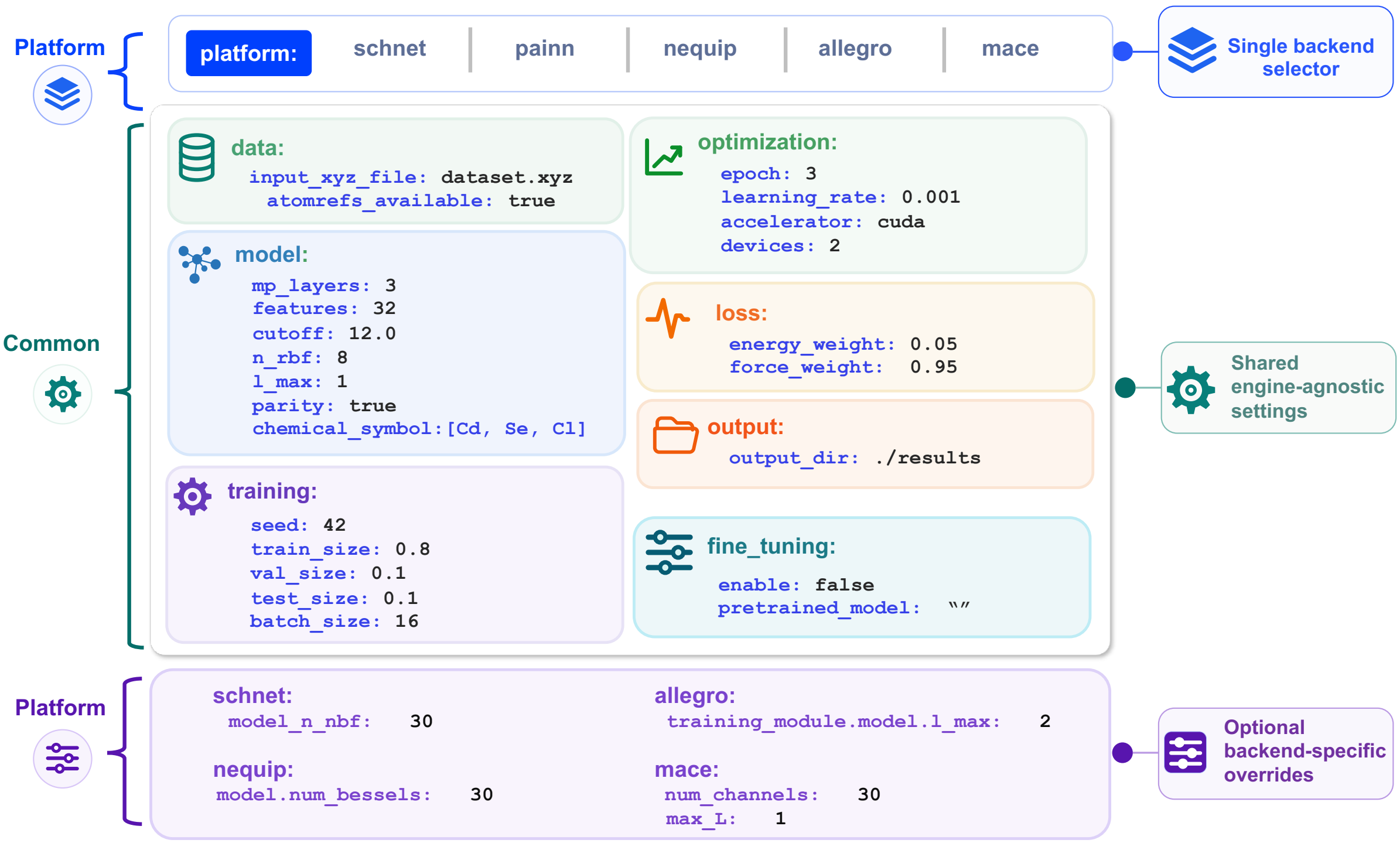

**Figure 3**: **Structure of the unified Orchestr.AI master YAML file.** The configuration consists of three main sections: platform selection, which specifies the MLIP engine; common settings, which define the dataset, model architecture, training parameters, optimization, loss function, output, and optional fine-tuning settings; and platform-specific settings, which allow expert users to define engine-specific parameters and overrides. The common settings are translated into native configurations for the selected engine. Additional SchNetPack architectures, including SO3net and FieldSchNet, are not illustrated.

In the unified mode, the experiment is defined through a master YAML file, whose structure is illustrated in **Figure 3**. The *common* block contains the quantities that should remain consistent across the selected engines, including the input dataset, chemical species, cutoff radius, model size, data partitions, random seed, number of training epochs, batch size, learning rate, loss weights, computational device settings, and output location. These parameters describe what should be trained and under which experimental conditions, without requiring the user to know where each native framework expects the corresponding option.

Orchestr.AI translates this common specification using explicit mappings associated with each engine. This translation is semantic rather than a direct renaming of YAML keys. For example, a shared radial-basis setting can be mapped to the radial basis functions used by SchNet or PaiNN and to the corresponding Bessel-basis parameter used by NequIP or Allegro. Similarly, common energy and force loss weights must be inserted into different parts of the native configuration hierarchy for each framework. The mapping layer therefore makes the intended correspondence between parameters explicit and repeatable, rather than leaving it to manually maintain input files whose settings may gradually diverge. When a shared parameter has no native architectural equivalent, such as spherical-harmonic degree (*l_max*) and parity in invariant SchNet or fixed-vector PaiNN, the adapter automatically omits the inapplicable key while mapping *mp_layers* to either message-passing interaction blocks (SchNet, PaiNN, NequIP, MACE) or local tensor-product layers (Allegro).

Not every native option can or should be reduced to a shared parameter. The master YAML consequently includes an optional overrides block in which advanced settings can be expressed using the native key hierarchy of a particular engine. These overrides provide access to architecture-specific functionality or targeted single-engine adjustments without expanding the common schema with parameters that have no meaningful equivalent elsewhere. By default, the common block establishes the shared baseline across engines, while explicit entries in the engine-specific override block can either supply unmapped native options or selectively supersede a mapped default when deliberately customizing an individual backend. Thus, the unified interface establishes comparable settings where equivalence is meaningful without forcing distinct architectures into an artificial lowest common denominator.

Before execution, Orchestr.AI generates a complete native YAML file for the selected engine and validates the associated dataset paths and input format. SchNet and PaiNN are executed through SchNetPack, while NequIP, Allegro, and MACE retain their respective native configuration structures and training procedures. The generated file can either be passed directly to training or produced independently for inspection and modification. This intermediate, human-readable configuration is important for transparency: users can verify exactly how the master specification has been interpreted before submitting a computationally expensive calculation, archive the effective native input together with the results, or use it as the starting point for a more specialized experiment.

Orchestr.AI also supports a native configuration mode for cases in which complete engine-specific control is required. In this mode, an existing native YAML file is supplied directly, while the platform continues to provide input validation, execution, environment dispatch, and standardized output handling. The two modes therefore address complementary use cases. Unified mode is appropriate for routine training and controlled cross-engine studies, whereas native mode allows experienced users to access parameters that cannot be represented through the common abstraction.

*2.3 Engine Dependencies and Unified Logs*

A further practical difficulty arises from the incompatible and rapidly evolving dependencies of the supported frameworks. Rather than installing every engine in a single software environment, Orchestr.AI associates each engine with an isolated environment profile. SchNetPack-based models, including SchNet and PaiNN, are assigned to the core environment; NequIP and Allegro share a separate NequIP environment; and MACE is executed in its own environment. The user launches the workflow through a common command, after which Orchestr.AI detects the requested engine and transfers execution to the appropriate environment. This design isolates conflicting dependencies while allowing each model to remain connected to its native and independently maintained software stack.

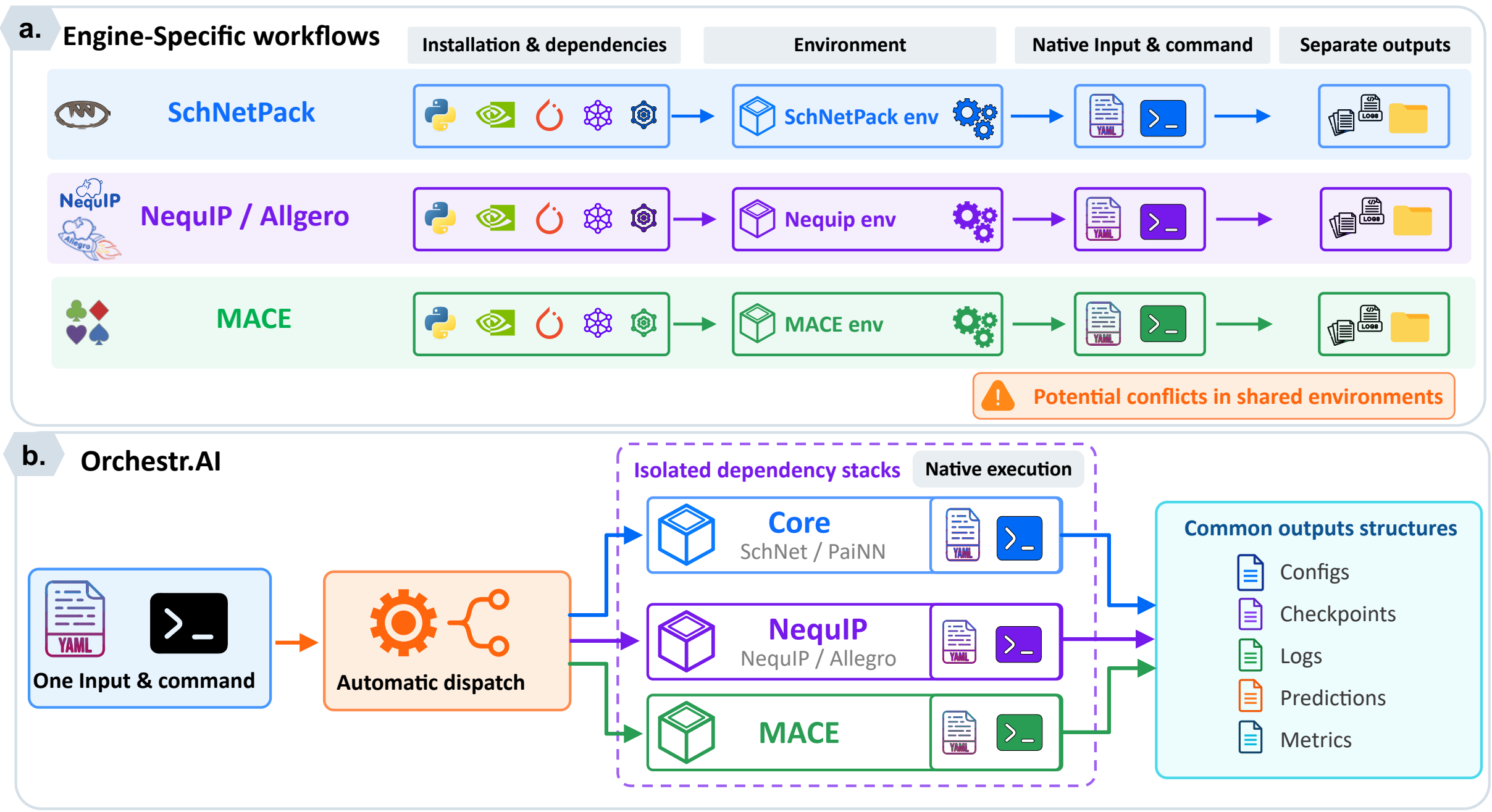

**Figure 4**: **Engine-specific dependencies and unified execution in Orchestr.AI. (a)** Conventional engine-specific workflows require separate dependency stacks, software environments, native configurations and commands, and output organization for SchNetPack, NequIP/Allegro, and MACE. **(b)** Orchestr.AI uses a common input and command to automatically dispatch execution to isolated engine environments while preserving native backend execution and collecting configurations, checkpoints, logs, predictions, and metrics within a standardized output structure.

**Figure 4** illustrates this organization by contrasting conventional engine-specific workflows (**Figure 4a**) with the Orchestr.AI approach (**Figure 4b**), in which a common input is automatically dispatched to isolated engine environments while preserving native execution. Hardware configuration is handled through the same experiment specification. The user declares the accelerator and number of devices, and Orchestr.AI translates these choices into the distributed-training settings expected by the selected backend. For example, the platform activates or removes the distributed data-parallel configuration for NequIP and Allegro according to the effective device count, while the execution wrappers apply the corresponding launch procedure for the other engines. The requested resources must still be allocated by the local scheduler, but the user no longer needs to reproduce the internal distributed-training syntax of every framework.

As additionally shown in **Figure 4b**, after training, Orchestr.AI collects the generated configurations, checkpoints, logs, predictions, and evaluation results within a common directory structure. Standardizing the organization of these outputs does not alter the files produced by the native engines; instead, it provides a predictable location and naming structure from which results can be inspected or compared.

*2.4 Post-training Simulations and Uncertainty Analysis*

Beyond model training and evaluation, Orchestr.AI provides a post-training environment in which the resulting machine-learning potential can be used directly for atomistic simulations (**Figure 5**). Through its interface with the Atomic Simulation Environment (ASE), trained models from any of the supported machine-learning engines can be deployed through a common workflow. As shown in **Figure 5a**, the current implementation supports geometry optimization, molecular dynamics with different integration and temperature-control schemes, and vibrational analysis[34]. This ASE-based design separates the simulation protocol from the underlying model architecture, allowing equivalent calculations to be performed without developing engine-specific post-processing workflows.

In **Figure 5b, we** show how trajectories generated through the same ASE interface can subsequently be evaluated using the batch-inference capabilities of Orchestr.AI. Rather than processing configurations individually, the trajectory is divided into configurable batches, enabling efficient prediction of energies and atomic forces and, where supported by the underlying backend (such as SchNetPack and MACE), extraction of latent model representations. The same procedure may also be applied to trajectories generated externally, providing a general route for screening large collections of molecular configurations with a trained potential.

**Figure 5c** illustrates how, when several independently trained models are available, Orchestr.AI can additionally use their prediction spread (quantified via the ensemble standard deviation of total energies and atomic force components) to estimate epistemic uncertainty in the energies and forces.[55] These uncertainty estimates can be calibrated against held-out validation residuals via a scaling factor and evaluated along an unlabeled trajectory [56]. Configurations associated with elevated uncertainty can then be filtered according to geometric and physical criteria (such as minimum interatomic distance thresholds to exclude unphysical atomic clashes) and selected for new quantum-chemical calculations. In this way, the post-training workflow can be extended naturally into an active-learning cycle, in which informative configurations are identified and used to improve the training dataset iteratively [57].

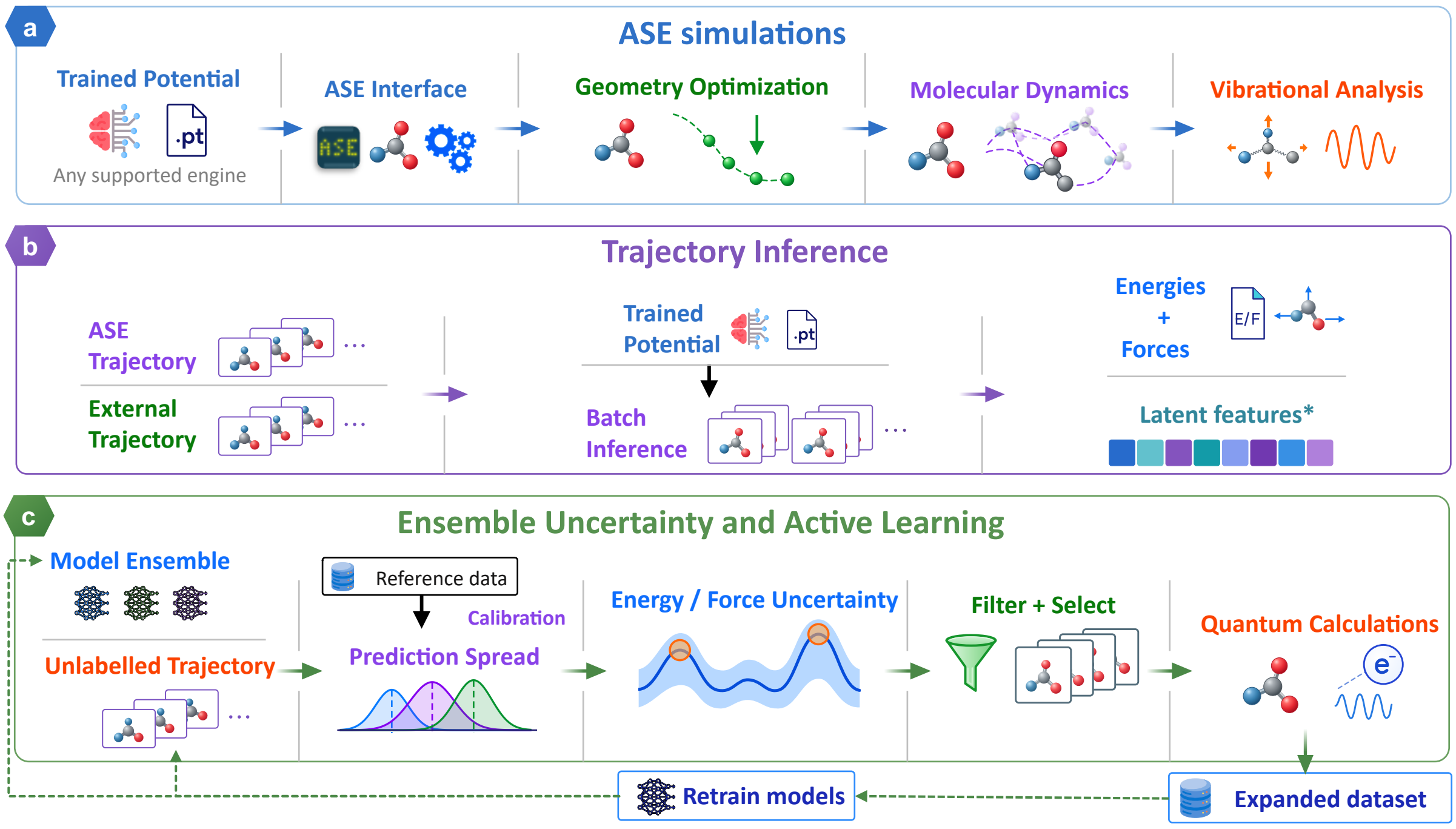

**Figure 5**: **Post-training simulations and uncertainty analysis in Orchestr.AI. (a)** Trained machine-learning interatomic potentials can be deployed through the ASE interface for geometry optimization, molecular dynamics, and vibrational analysis. **(b)** Trajectories generated through ASE or provided externally can be processed by batch inference to obtain energies, forces, and, where supported, latent representations. **(c)** When multiple independently trained models are available, prediction spread can be used to estimate and calibrate epistemic uncertainty, filter informative configurations, and support active-learning cycles through new quantum-chemical calculations and iterative retraining.

## 3. Results and Benchmark

A unified interface is useful only if it can coordinate different engines while still producing results that can be compared meaningfully. **Figure 6** presents the automated benchmarking workflow in Orchestr.AI, which proceeds through five stages: (1) input setup of the dataset, master YAML, and HPC settings; (2) benchmark-space expansion into engine- and hyperparameter-specific subfolders; (3) validated native YAML and SLURM script generation; (4) parallel multi-node GPU execution via the job scheduler; and (5) automated result collection and metric extraction across multiple MLFF engines. We test this by using one Orchestr.AI workflow to prepare the data, generate native configurations, train SchNet, PaiNN, NequIP, Allegro, and MACE, and evaluate their predictions [58]. Common data partitions and evaluation metrics provide the basis for comparison, while each model retains its native implementation. We first describe the benchmark setup and cross-engine harmonization strategy, then contextualize the workflow against previously reported results before examining performance on rMD17 and 3BPA and assessing repeated-run reproducibility.

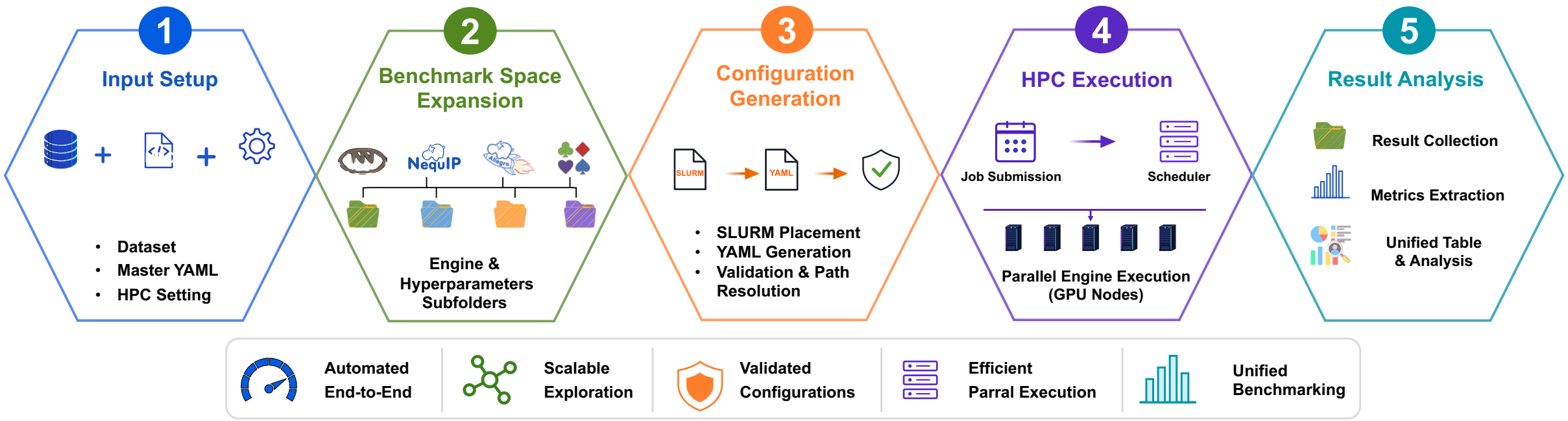

**Figure 6: Automated benchmarking workflow in Orchestr.AI**. The workflow comprises five stages: (1) input setup, including the dataset, YAML configuration, and high-performance computing (HPC) settings; (2) benchmark-space expansion into engine- and hyperparameter-specific subfolders; (3) generation and validation of native configurations and execution scripts; (4) parallel execution of benchmark jobs on HPC resources; and (5) automated result collection, metric extraction, and comparative analysis across MLIP engines.

### *3.1. Benchmark setup*

To assess the consistency and reproducibility of Orchestr.AI across different MLIP engines, we trained five widely used architectures—SchNet, PaiNN, NequIP, Allegro, and MACE—using the unified Orchestr.AI training pipeline.

All models were trained on the rMD17 dataset [59] , which contains DFT molecular-dynamics trajectories for small organic molecules. The molecular configurations were extracted from the original NPZ files using the five predefined training and test splits provided with the dataset, each specified by separate CSV index files. For each split, 1,000 configurations were selected for model development, comprising 950 for training and 50 for validation, while an independent set of 1,000 configurations was used for testing. The procedure was repeated across five independent data splits to reduce the influence of sampling variability. Performance is reported as the mean absolute error (MAE) in energies (meV) and forces (meV/Å), averaged over the five runs. In addition to the rMD17 benchmark, model generalization was evaluated using the 3BPA dataset[60], trains on 500 configurations at 300 K and includes independent test configurations sampled at different temperatures (300, 600, and 1200 K) as well as structures exhibiting different dihedral distortions.

The benchmarks presented here evaluate current, actively maintained implementations of the supported MLIP engines within a unified workflow, rather than attempting to reproduce the exact software environments used in the original architecture studies. In Orchestr.AI version 0.2.0 [61], reproducibility is achieved through fixed engine profiles instead of a single shared software environment. Each engine is executed within its corresponding profile, allowing mutually compatible and maintained software versions to be used without forcing all frameworks into the same environment. For the calculations reported here, these profiles included SchNetPack 2.2.0, NequIP 0.17.1, Allegro 0.8.2, and MACE 0.3.16.

The objective is therefore to compare the behavior of the different architectures under consistent data processing, environment dispatch, and training and evaluation protocols, rather than to reproduce the precise numerical values reported in earlier engine-specific studies. The use of identical data splits and a common preprocessing and configuration workflow ensure a controlled comparison of the supported engines within Orchestr.AI.

### *3.2. Cross-engine benchmark harmonization and relation to published baselines*

All benchmarks presented in this work were performed using engine-specific YAML configurations, documented in the accompanying input-files repository. While Orchestr.AI supports a unified master YAML interface for cross-engine configuration mapping, as illustrated in **Figure 3**, the present benchmarks employed engine-specific configurations based on the model-specific settings reported in the corresponding reference studies, retaining their principal architectural and training characteristics rather than imposing identical hyperparameters across architectures. Orchestr.AI coordinated the execution of SchNet, PaiNN, NequIP, Allegro, and MACE through their respective native software environments while maintaining consistent dataset partitions, evaluation procedures, and output organization [6–8,12–15]. This approach provides a common framework for cross-engine benchmarking while preserving the distinct model configurations and training procedures associated with each architecture. Workflow reproducibility was assessed by repeating the five-split benchmarks three times using the same engine-specific configurations and fixed software environment profiles.

To contextualize the benchmark results obtained with Orchestr.AI, we compared them with the values reported in the original publications describing SchNet, PaiNN, NequIP, Allegro, and MACE [6–8,12–15]. While published hyperparameters were translated as closely as possible into the corresponding settings available in the maintained APIs, exact reproduction of the original training setups is not always feasible because the underlying frameworks and software dependencies have continued to evolve. Several of these engines rely on rapidly developing libraries, including e3nn, in which successive releases have introduced changes to model definitions, configuration interfaces, and available hyperparameters. For example, earlier implementations permitted the dimensions of individual neural-network layers to be specified explicitly, whereas current interfaces may describe network depth and width through a smaller set of higher-level parameters. Consequently, the models evaluated here retain the essential characteristics of the published architectures but are not necessarily identical to their original software implementations.

The goal of this comparison is therefore to determine whether the expected qualitative performance trends are recovered within a unified and reproducible workflow. As shown in **Tables 1** and **Table 2**, the absolute errors differ from some previously reported values, but the overall trends remain consistent: equivariant architectures such as NequIP and Allegro generally outperform earlier invariant models, while MACE provides competitive accuracy for both energy and force predictions. These results indicate that Orchestr.AI preserves the characteristic performance of the supported architectures while enabling their evaluation through a common workflow.

### *3.3. Benchmark results on rMD17 and 3BPA*

For the rMD17 benchmark, the platform prepares the same data partitions, translates a shared experiment specification into each engine's required inputs, and collects predictions for evaluation using the same metrics. **Table 1** reports the resulting errors for aspirin, benzene, ethanol, paracetamol, and uracil. NequIP achieves the lowest or joint-lowest energy error for every molecule, whereas the best force result varies among NequIP, Allegro, and MACE.

Aspirin makes this distinction particularly clear. SchNet has the largest energy and force errors, at 15.73 meV and 40.15 meV/Å, respectively. NequIP reduces the energy error to 0.12 meV, but MACE provides the lowest force error at 7.98 meV/Å; NequIP and Allegro follow at 9.03 and 9.46 meV/Å. Thus, even among the strongest models, the preferred architecture depends on which property is most important for the intended application.

**Table 1.** Mean absolute errors obtained for the rMD17 dataset using the five MLFF architectures. Energy (E) and force (F) errors are reported in meV and meV/Å, respectively. For each molecule, the models were trained using 950 configurations and validated against 50 configurations.

| | | SchNet | PaiNN | Nequip | Allegro | MACE |
|---|---|---|---|---|---|---|
| **Aspirin** | E | 15.73 | 9.26 | 0.12 | 0.24 | 4.97 |
| | F | 40.15 | 15.54 | 9.03 | 9.46 | 7.98 |
| **Benzene** | E | 2.16 | 0.33 | 0.01 | 0.01 | 1.2 |
| | F | 10.33 | 1.25 | 0.61 | 0.56 | 0.6 |
| **Ethanol** | E | 2.77 | 2.96 | 0.05 | 0.07 | 2.88 |
| | F | 13.42 | 8.23 | 3.31 | 3.47 | 3.58 |
| **Paracetamol** | E | 9.69 | 6.20 | 0.07 | 0.47 | 4.8 |
| | F | 30.02 | 11.39 | 5.90 | 6.59 | 5.72 |
| Uracil | E | 4.22 | 1.65 | 0.04 | 0.08 | 7.68 |
| | F | 19.64 | 6.13 | 3.65 | 3.18 | 3.96 |

Benzene provides a contrast: the leading models are much more closely grouped. NequIP and Allegro both have an energy error of 0.01 meV at the reported precision, while the force errors of Allegro, MACE, and NequIP range from 0.56 to 0.61 meV/Å. For ethanol, NequIP leads on both properties. The distinction reappears for paracetamol and uracil, where NequIP gives the lowest energy errors, but MACE and Allegro, respectively, give the lowest force errors.

These results illustrate the practical value of the unified workflow. The energy–force trade-offs can be examined using the same molecular partitions and evaluation procedure, with each engine's effective configuration and outputs retained for inspection. Orchestr.AI does not make the architectures equivalent; it makes the conditions of their comparison more explicit and repeatable. The broad performance trends agree with previous reports [6–8,12–15] , while the differences across molecules and properties show why no single architecture should be selected from one error metric alone.

*3BPA generalization:*

To examine model robustness beyond the configurations represented during training, models trained exclusively on structures sampled at 300 K were evaluated against independent configurations at 300, 600, and 1200 K. They were also tested on dihedral slices representing controlled torsional distortions. The temperature series progressively increases the configurational diversity, while the dihedral slices provide a complementary test of generalization along specific conformational degrees of freedom. The resulting errors are summarized in **Table 2.**

**Table 2.** Root-mean-square errors (RMSEs) obtained for the 3BPA dataset using models trained on configurations sampled at 300 K. Energy (E) and force (F) errors are reported in meV and meV/Å, respectively, for independent configurations sampled at 300, 600, and 1200 K and for the dihedral-slice dataset.

| | | SchNet | PaiNN | NequIP | Allegro | MACE |
|---|---|---|---|---|---|---|
| 300 K | E | 17.16 | 9.49 | 3.5 | 4.45 | 5.4 |
| | F | 50.65 | 24.49 | 11.69 | 14.30 | 10.6 |
| 600 K | E | 55.25 | 34.66 | 12.35 | 12.98 | 13.5 |
| | F | 119.74 | 75.68 | 28.60 | 33.22 | 28.2 |
| 1200 K | E | 190.84 | 236.62 | 38.44 | 54.65 | 48.6 |
| | F | 373.0 | 354.98 | 84.17 | 103.03 | 101.1 |
| | E | 64.26 | 34.45 | 18.27 | 16.69 | 32.4 |
| **Dihedral Slices** | F | 85.45 | 46.18 | 22.63 | 26.48 | 27.2 |

For the independent 300 K test set, all models retain relatively low errors. NequIP achieves the lowest energy RMSE of 3.50 meV, while MACE produces the lowest force RMSE of 10.60 meV/Å, closely followed by NequIP at 11.69 meV/Å. Allegro also performs competitively, whereas SchNet and PaiNN show appreciably larger force errors, indicating a reduced ability to generalize to configurations not encountered during training [13,14].

As the temperature increases, the sampled structures extend progressively further from the 300 K training distribution, making the prediction task more demanding. At 600 K, the force errors increase for every model, reaching 119.74 meV/Å for SchNet, whereas MACE and NequIP maintain substantially lower errors of 28.20 and 28.60 meV/Å, respectively. This trend becomes more pronounced at 1200 K, where larger configurational diversity leads to a further increase in both energy and force errors. Nevertheless, NequIP, Allegro, and MACE retain a clear advantage over SchNet and PaiNN, showing that the relative performance trends remain broadly consistent under increasingly challenging conditions.

The dihedral-slice test provides a complementary assessment by probing torsional distortions that may not be adequately represented in the training data. Allegro achieves the lowest energy RMSE of 16.69 meV, while NequIP gives the lowest force RMSE of 22.63 meV/Å and remains competitive for both properties. PaiNN and MACE show intermediate performance, whereas SchNet again produces the largest errors. These results demonstrate that the differences among the architectures persist not only with increasing thermal diversity but also for targeted out-of-distribution conformational changes.

*Repeated-run reproducibility.*

For the rMD17 benchmark, the result for each molecule–engine combination was first averaged over the five predefined data splits. The mean and standard deviation were then calculated across the three repeated benchmark executions, as reported in **Table 3**. Equivalent repeated-run experiments were performed for the 3BPA benchmark and produced the same overall conclusions regarding workflow stability; for conciseness, only the rMD17 results are presented here. The repeated executions produce highly consistent MAEs across different engines. For SchNet, PaiNN, and NequIP, most energy and force standard deviations are zero at the reported precision or remain very small, indicating that the same YAML configuration and fixed engine profile reproduce nearly identical results. Allegro shows modest residual variation, while MACE exhibits greater variability for some energy predictions. Force errors nevertheless remain comparatively stable across the investigated systems. Such residual differences are expected in deep-learning training because of backend-level numerical , GPU operations, data loading, and engine-specific optimization dynamics. The standard deviations should therefore be interpreted as evidence of practical workflow reproducibility rather than bitwise-deterministic training.

These results show that Orchestr.AI provides repeatable benchmark execution and cross-engine compatibility when software versions, dependencies, engine profiles, and input configurations are fixed. By standardizing experiment definition, dataset partitioning, environment dispatch, and training execution, the platform enables transparent and reproducible comparisons across multiple MLFF engines.

**Table 3.** Repeated-run stability of Orchestr.AI for the rMD17 benchmark. Energy (E) and force (F) MAEs are reported as the mean ± standard deviation over three repeated benchmark executions. Within each execution, the result for every molecule–engine combination was first averaged over the five predefined rMD17 data splits.

| | | SchNet | PaiNN | Nequip | Allegro | MACE |
|---|---|---|---|---|---|---|
| Aspirin | E | 15.727 ± 0.000 | 9.258 ± 0.000 | 0.124 ± 0.000 | 0.247 ± 0.023 | 4.480 ± 0.485 |
| | F | 40.150 ± 0.000 | 15.540 ± 0.000 | 9.039 ± 0.000 | 9.412 ± 0.043 | 7.993 ± 0.023 |
| Benzene | E | 2.159 ± 0.000 | 0.326 ± 0.000 | 0.006 ± 0.000 | 0.017 ± 0.009 | 1.200 ± 0.240 |
| | F | 10.325 ± 0.000 | 1.254 ± 0.000 | 0.614 ± 0.003 | 0.559 ± 0.003 | 0.607 ± 0.012 |
| Ethanol | E | 2.769 ± 0.000 | 2.963 ± 0.000 | 0.055 ± 0.000 | 0.079 ± 0.012 | 2.220 ± 0.632 |
| | F | 13.421 ± 0.000 | 8.231 ± 0.000 | 3.309 ± 0.001 | 3.445 ± 0.037 | 3.620 ± 0.040 |
| Paracetamol | E | 9.687 ± 0.000 | 6.195 ± 0.000 | 0.071 ± 0.000 | 0.604 ± 0.112 | 6.133 ± 2.663 |
| | F | 30.019 ± 0.000 | 11.388 ± 0.000 | 5.895 ± 0.000 | 6.600 ± 0.027 | 5.740 ± 0.020 |
| Uracil | E | 4.224 ± 0.000 | 1.648 ± 0.000 | 0.043 ± 0.000 | 0.142 ± 0.107 | 7.360 ± 1.466 |
| | F | 19.636 ± 0.000 | 6.128 ± 0.000 | 3.648 ± 0.001 | 3.244 ± 0.058 | 3.987 ± 0.046 |

## 4. Discussion and Conclusions

The benchmarks presented in this work demonstrate that Orchestr.AI can consistently train and evaluate multiple machine-learning force-field architectures through a common workflow. Across the rMD17 systems, NequIP generally produces the lowest energy errors, whereas the best force accuracy is distributed among NequIP, Allegro, and MACE depending on the molecule. On the more demanding 3BPA benchmark, errors increase with thermal and torsional diversity, yet NequIP, Allegro, and MACE retain a clear advantage over invariant and lower-order models. Rather than establishing a universal ranking of architectures, these benchmarks show that Orchestr.AI reproduces expected qualitative trends under controlled data-processing and training protocols. In particular, they highlight that the model yielding the lowest energy error does not necessarily provide the most accurate forces—and that imposing a uniform single-stage loss schedule can affect convergence differently across engines, as seen in the elevated energy variance of MACE relative to its stable force predictions. Exposing such trade-offs under transparent conditions is a primary advantage of the platform.

Orchestr.AI addresses an infrastructure problem rather than introducing a new machine-learning architecture. By separating the scientific definition of an experiment in a master YAML configuration from its engine-specific implementation, the platform eliminates the need to repeatedly adapt datasets, configuration files, and computing environments across fragmented software ecosystems. Pairing this declarative abstraction with isolated, fixed engine profiles avoids dependency conflicts among rapidly evolving libraries while preserving native model execution. Consequently, even though evolving framework APIs preclude exact bitwise reproduction of historical publications, Orchestr.AI achieves high repeated-run stability and transparent protocol reproducibility using actively maintained software stacks.

Beyond training and static benchmarking, Orchestr.AI provides an extensible end-to-end architecture that integrates dataset consolidation, representative-structure selection, fine-tuning, ASE-based simulations, batch inference, and ensemble-driven epistemic uncertainty calibration for active learning. While the present empirical benchmarks focus on molecular accuracy and workflow repeatability across rMD17 and 3BPA, broader validation will require larger, chemically diverse, and periodic materials datasets, alongside systematic profiling of training runtimes, memory footprints, and active-learning loops across model families.

Future developments will expand support to additional MLFF engines, automated hyperparameter exploration, and large-scale high-performance-computing workflows. In conclusion, by combining engine-independent experiment definitions with isolated software environments, Orchestr.AI reduces the technical overhead of cross-framework experimentation and provides a reproducible foundation for developing, comparing, and deploying machine-learning interatomic potentials.

## ASSOCIATED CONTENT

The source code of Orchestr.AI is publicly available on GitHub at **https://github.com/nlesc-nano/Orchestr.AI** under the GPL-3.0 license; the results reported in this work were obtained with version 0.2.0, archived on Zenodo [61]. The engine-specific YAML configuration files and associated input files used for all benchmarks are available at **https://github.com/nlesc-**

**nano/input-files**, enabling the reported experiments to be reproduced. The rMD17 [59] and 3BPA [60] datasets are publicly available from their original sources. Raw benchmark outputs and trained models are available from the corresponding author upon request.

AUTHOR INFORMATION

**Corresponding Author**

*ivan.infante@bcmaterials.net

ORCID: 0000-0003-3467-9376

**Funding Sources**

This research received funding from the European Union's Horizon Europe programme through the EIC Pathfinder project UNICORN (Grant Agreement No. 101098649) and the Marie Skłodowska-Curie Actions Doctoral Network Track The Twin (Grant Agreement No. 101168820). Additional support was provided through the IKUR Strategy under the collaboration agreement between the Ikerbasque Foundation and BCMaterials, on behalf of the Department of Education of the Basque Government, and through grants supporting the activities of research groups of the Basque university system 2026–2029 (IT2001-26).

The authors acknowledge the technical and human support provided by the DIPC Supercomputing Center and the computational resources made available through the Hyperion cluster at the Donostia International Physics Center (DIPC). Additionally, the authors gratefully acknowledge the EuroHPC Joint Undertaking for awarding access to the LEONARDO supercomputer, hosted by CINECA (Italy) and the LEONARDO consortium, through a EuroHPC Access call.

**Notes**

The author declares no competing financial interest.

## References

(1) Chmiela, S.; Tkatchenko, A.; Sauceda, H. E.; Poltavsky, I.; Schütt, K. T.; Müller, K.-R. Machine Learning of Accurate Energy-Conserving Molecular Force Fields. *Sci. Adv.* **2017**, *3* (5), e1603015. https://doi.org/10.1126/sciadv.1603015

(2) Unke, O. T.; Chmiela, S.; Sauceda, H. E.; Gastegger, M.; Poltavsky, I.; Schütt, K. T.; Tkatchenko, A.; Müller, K. R. Machine Learning Force Fields. *Chem. Rev.* **2021**, *121* (16), 10142–10186. https://doi.org/10.1021/ACS.CHEMREV.0C01111

(3) Bartók, A. P.; Payne, M. C.; Kondor, R.; Csányi, G. Gaussian Approximation Potentials: The Accuracy of Quantum Mechanics, without the Electrons. *Phys. Rev. Lett.* **2010**, *104* (13), 136403. https://doi.org/10.1103/PhysRevLett.104.136403

(4) S. Smith, J.; Isayev, O.; E. Roitberg, A. ANI-1: An Extensible Neural Network Potential with DFT Accuracy at Force Field Computational Cost. *Chem. Sci.* **2017**, *8* (4), 3192–3203. https://doi.org/10.1039/C6SC05720A

(5) Kim, J.; Cho, H.; Jeon, H.; Jung, J.; Han, S. Reactive Machine Learning Interatomic Potentials for Chemistry and Materials Science. *Chem. Rev.* **2026**, *126* (8), 4467–4510. https://doi.org/10.1021/ACS.CHEMREV.5C00728

(6) Schütt, K. T.; Kessel, P.; Gastegger, M.; Nicoli, K. A.; Tkatchenko, A.; Müller, K.-R. SchNetPack: A Deep Learning Toolbox For Atomistic Systems. *J. Chem. Theory Comput.* **2019**, *15* (1), 448–455. https://doi.org/10.1021/acs.jctc.8b00908

(7) Schütt, K. T.; Hessmann, S. S. P.; Gebauer, N. W. A.; Lederer, J.; Gastegger, M. SchNetPack 2.0: A Neural Network Toolbox for Atomistic Machine Learning. *J. Chem. Phys.* **2023**, *158* (14), 144801. https://doi.org/10.1063/5.0138367

(8) Schütt, K.; Kindermans, P.-J.; Sauceda Felix, H. E.; Chmiela, S.; Tkatchenko, A.; Müller, K.-R. SchNet: A Continuous-Filter Convolutional Neural Network for Modeling Quantum Interactions; Curran Associates, Inc., 2017; Vol. 30.

(9) Gasteiger, J.; Groß, J.; Günnemann, S. Directional Message Passing for Molecular Graphs. *8th International Conference on Learning Representations, ICLR 2020* **2020**.

(10) Gasteiger, J.; Becker, F.; Günnemann, S. GemNet: Universal Directional Graph Neural Networks for Molecules. *Adv. Neural Inf. Process. Syst.* **2021**, *34*, 6790–6802.

(11) Anstine, D. M.; Zubatyuk, R.; Isayev, O. AIMNet2: A Neural Network Potential to Meet Your Neutral, Charged, Organic, and Elemental-Organic Needs. *Chem. Sci.* **2025**, *16* (23), 10228–10244. https://doi.org/10.1039/D4SC08572H

(12) Schütt, K.; Unke, O.; Gastegger, M. Equivariant Message Passing for the Prediction of Tensorial Properties and Molecular Spectra; PMLR, 2021; pp 9377–9388.

(13) Musaelian, A.; Batzner, S.; Johansson, A.; Sun, L.; Owen, C. J.; Kornbluth, M.; Kozinsky, B. Learning Local Equivariant Representations for Large-Scale Atomistic Dynamics. *Nat. Commun.* **2023**, *14* (1), 579. https://doi.org/10.1038/s41467-023-36329-y

(14) Batzner, S.; Musaelian, A.; Sun, L.; Geiger, M.; Mailoa, J. P.; Kornbluth, M.; Molinari, N.; Smidt, T. E.; Kozinsky, B. E(3)-Equivariant Graph Neural Networks for Data-Efficient and Accurate Interatomic Potentials. *Nat. Commun.* **2022**, *13* (1), 2453. https://doi.org/10.1038/s41467-022-29939-5

(15) Batatia, I.; Kovacs, D. P.; Simm, G.; Ortner, C.; Csányi, G. MACE: Higher Order Equivariant Message Passing Neural Networks for Fast and Accurate Force Fields. *Adv. Neural Inf. Process. Syst.* **2022**, *35*, 11423–11436.

(16) Frank, J. T.; Unke, O. T.; Müller, K. R.; Chmiela, S. A Euclidean Transformer for Fast and Stable Machine Learned Force Fields. *Nature Communications* **2024**, *15* (1), 6539-. https://doi.org/10.1038/s41467-024-50620-6

(17) Simeon, G.; De Fabritiis, G. TensorNet: Cartesian Tensor Representations for Efficient Learning of Molecular Potentials. *Adv. Neural Inf. Process. Syst.* **2023**, *36*, 37334–37353. https://doi.org/10.52202/075280-1623
(18) Fu, X.; Wood, B. M.; Barroso-Luque, L.; Levine, D. S.; Gao, M.; Dzamba, M.; Zitnick, C. L. Learning Smooth and Expressive Interatomic Potentials for Physical Property Prediction. PMLR October 6, 2025, pp 17875–17893. https://proceedings.mlr.press/v267/fu25h.html (accessed 2026-09-24)
(19) Park, Y.; Kim, J.; Hwang, S.; Han, S. Scalable Parallel Algorithm for Graph Neural Network Interatomic Potentials in Molecular Dynamics Simulations. *J. Chem. Theory Comput.* **2024**, *20* (11), 4857–4868. https://doi.org/10.1021/ACS.JCTC.4C00190
(20) Liao, Y.-L.; Wood, B.; Das, A.; Smidt, T. EquiformerV2: Improved Equivariant Transformer for Scaling to Higher-Degree Representations. *International Conference on Learning Representations* **2024**, *2024*, 39282–39309.
(21) Chen, C.; Ong, S. P. A Universal Graph Deep Learning Interatomic Potential for the Periodic Table. *Nature Computational Science* **2022**, *2* (11), 718–728. https://doi.org/10.1038/s43588-022-00349-3
(22) Deng, B.; Zhong, P.; Jun, K. J.; Riebesell, J.; Han, K.; Bartel, C. J.; Ceder, G. CHGNet as a Pretrained Universal Neural Network Potential for Charge-Informed Atomistic Modelling. *Nature Machine Intelligence* **2023**, *5* (9), 1031–1041. https://doi.org/10.1038/s42256-023-00716-3
(23) Batatia, I.; Benner, P.; Chiang, Y.; Elena, A. M.; Kovács, D. P.; Riebesell, J.; Advincula, X. R.; Asta, M.; Avaylon, M.; Baldwin, W. J.; Berger, F.; Bernstein, N.; Bhowmik, A.; Bigi, F.; Blau, S. M.; Cărare, V.; Ceriotti, M.; Chong, S.; Darby, J. P.; De, S.; Della Pia, F.; Deringer, V. L.; Elijošius, R.; El-Machachi, Z.; Fako, E.; Falcioni, F.; Ferrari, A. C.; Gardner, J. L. A.; Gawkowski, M. J.; Genreith-Schriever, A.; George, J.; Goodall, R. E. A.; Grandel, J.; Grey, C. P.; Grigorev, P.; Han, S.; Handley, W.; Heenen, H. H.; Hermansson, K.; Ho, C. H.; Hofmann, S.; Holm, C.; Jaafar, J.; Jakob, K. S.; Jung, H.; Kapil, V.; Kaplan, A. D.; Karimitari, N.; Kermode, J. R.; Kourtis, P.; Kroupa, N.; Kullgren, J.; Kuner, M. C.; Kuryla, D.; Liepuoniute, G.; Lin, C.; Margraf, J. T.; Magdău, I.-B.; Michaelides, A.; Moore, J. H.; Naik, A. A.; Niblett, S. P.; Norwood, S. W.; O'Neill, N.; Ortner, C.; Persson, K. A.; Reuter, K.; Rosen, A. S.; Rosset, L. A. M.; Schaaf, L. L.; Schran, C.; Shi, B. X.; Sivonxay, E.; Stenczel, T. K.; Sutton, C.; Svahn, V.; Swinburne, T. D.; Tilly, J.; van der Oord, C.; Vargas, S.; Varga-Umbrich, E.; Vegge, T.; Vondrák, M.; Wang, Y.; Witt, W. C.; Wolf, T.; Zills, F.; Csányi, G. A Foundation Model for Atomistic Materials Chemistry. *J. Chem. Phys.* **2025**, *163* (18), 184110. https://doi.org/10.1063/5.0297006
(24) Lysogorskiy, Y.; Bochkarev, A.; Drautz, R. Graph Atomic Cluster Expansion for Foundational Machine Learning Interatomic Potentials. *npj Computational Materials* **2026**, *12* (1), 114-. https://doi.org/10.1038/s41524-026-01979-1
(25) Mazitov, A.; Bigi, F.; Kellner, M.; Pegolo, P.; Tisi, D.; Fraux, G.; Pozdnyakov, S.; Loche, P.; Ceriotti, M. PET-MAD as a Lightweight Universal Interatomic Potential for Advanced Materials Modeling. *Nat. Commun.* **2025**, *16* (1), 10653. https://doi.org/10.1038/s41467-025-65662-7
(26) Zhu, A.; Batzner, S.; Musaelian, A.; Kozinsky, B. Fast Uncertainty Estimates in Deep Learning Interatomic Potentials. *J. Chem. Phys.* **2023**, *158* (16), 164111. https://doi.org/10.1063/5.0136574
(27) Batatia, I.; Batzner, S.; Kovács, D. P.; Musaelian, A.; Simm, G. N. C.; Drautz, R.; Ortner, C.; Kozinsky, B.; Csányi, G. The Design Space of E(3)-Equivariant Atom-Centred Interatomic Potentials. *Nat. Mach. Intell.* **2025**, *7* (1), 56–67. https://doi.org/10.1038/s42256-024-00956-x

(28) Pelaez, R. P.; Simeon, G.; Galvelis, R.; Mirarchi, A.; Eastman, P.; Doerr, S.; Thölke, P.; Markland, T. E.; De Fabritiis, G. TorchMD-Net 2.0: Fast Neural Network Potentials for Molecular Simulations. *J. Chem. Theory Comput.* **2024**, *20* (10), 4076–4087. https://doi.org/10.1021/ACS.JCTC.4C00253

(29) Ko, T. W.; Deng, B.; Nassar, M.; Barroso-Luque, L.; Liu, R.; Qi, J.; Thakur, A. C.; Mishra, A. R.; Liu, E.; Ceder, G.; Miret, S.; Ong, S. P. Materials Graph Library (MatGL), an Open-Source Graph Deep Learning Library for Materials Science and Chemistry. *npj Computational Materials* **2025**, *11* (1), 253-. https://doi.org/10.1038/s41524-025-01742-y

(30) Geiger, M.; Smidt, T. E3nn: Euclidean Neural Networks. **2022**.

(31) Benedini, G.; Loew, A.; Hellström, M.; A Gardner, J. L.; Baker, K. T.; Deringer, V. L.; Leimeroth, N.; Erhard, L. C.; Albe, K.; Rohrer, J. Machine-Learning Interatomic Potentials from a Users Perspective: A Comparison of Accuracy, Speed and Data Efficiency. *Model. Simul. Mat. Sci. Eng.* **2025**, *33* (6), 065012. https://doi.org/10.1088/1361-651X/ADF56D

(32) Jacobs, R.; Morgan, D.; Attarian, S.; Meng, J.; Shen, C.; Wu, Z.; Xie, C. Y.; Yang, J. H.; Artrith, N.; Blaiszik, B.; Ceder, G.; Choudhary, K.; Csanyi, G.; Cubuk, E. D.; Deng, B.; Drautz, R.; Fu, X.; Godwin, J.; Honavar, V.; Isayev, O.; Johansson, A.; Kozinsky, B.; Martiniani, S.; Ong, S. P.; Poltavsky, I.; Schmidt, K.; Takamoto, S.; Thompson, A. P.; Westermayr, J.; Wood, B. M. A Practical Guide to Machine Learning Interatomic Potentials – Status and Future. *Curr. Opin. Solid State Mater. Sci.* **2025**, *35*, 101214. https://doi.org/10.1016/j.cossms.2025.101214

(33) Aldossary, A.; Campos-Gonzalez-Angulo, J. A.; Pablo-García, S.; Leong, S. X.; Rajaonson, E. M.; Thiede, L.; Tom, G.; Wang, A.; Avagliano, D.; Aspuru-Guzik, A. In Silico Chemical Experiments in the Age of AI: From Quantum Chemistry to Machine Learning and Back. *Advanced Materials* **2024**, *36* (30), 2402369. https://doi.org/10.1002/ADMA.202402369

(34) Hjorth Larsen, A.; JØrgen Mortensen, J.; Blomqvist, J.; Castelli, I. E.; Christensen, R.; Dułak, M.; Friis, J.; Groves, M. N.; Hammer, B.; Hargus, C.; Hermes, E. D.; Jennings, P. C.; Bjerre Jensen, P.; Kermode, J.; Kitchin, J. R.; Leonhard Kolsbjerg, E.; Kubal, J.; Kaasbjerg, K.; Lysgaard, S.; Bergmann Maronsson, J.; Maxson, T.; Olsen, T.; Pastewka, L.; Peterson, A.; Rostgaard, C.; SchiØtz, J.; Schütt, O.; Strange, M.; Thygesen, K. S.; Vegge, T.; Vilhelmsen, L.; Walter, M.; Zeng, Z.; Jacobsen, K. W. The Atomic Simulation Environment—a Python Library for Working with Atoms. *Journal of Physics: Condensed Matter* **2017**, *29* (27), 273002. https://doi.org/10.1088/1361-648X/AA680E

(35) Wen, M.; Afshar, Y.; Elliott, R. S.; Tadmor, E. B. KLIFF: A Framework to Develop Physics-Based and Machine Learning Interatomic Potentials. *Comput. Phys. Commun.* **2022**, *272*, 108218. https://doi.org/10.1016/J.CPC.2021.108218

(36) *OpenKIM · Knowledgebase of Interatomic Models · Interatomic Potentials and Force Fields*. https://openkim.org/kim-api/ (accessed 2026-09-24)

(37) Verma, R.; Joshi, N.; Pfaendtner, J. SPARC: An Automated Workflow Toolkit for Accelerated Active Learning of Reactive Machine Learning Interatomic Potentials. *J. Open Source Softw.* **2026**, *11* (121), 9468. https://doi.org/10.21105/JOSS.09468

(38) Zeng, J.; Zhang, D.; Peng, A.; Zhang, X.; He, S.; Wang, Y.; Liu, X.; Bi, H.; Li, Y.; Cai, C.; Zhang, C.; Du, Y.; Zhu, J. X.; Mo, P.; Huang, Z.; Zeng, Q.; Shi, S.; Qin, X.; Yu, Z.; Luo, C.; Ding, Y.; Liu, Y. P.; Shi, R.; Wang, Z.; Bore, S. L.; Chang, J.; Deng, Z.; Ding, Z.; Han, S.; Jiang, W.; Ke, G.; Liu, Z.; Lu, D.; Muraoka, K.; Oliaei, H.; Singh, A. K.; Que, H.; Xu, W.; Xu, Z.; Zhuang, Y. Bin; Dai, J.; Giese, T. J.; Jia, W.; Xu, B.; York, D. M.; Zhang, L.; Wang, H. DeePMD-Kit v3: A Multiple-Backend Framework for Machine

Learning Potentials. *J. Chem. Theory Comput.* **2025**, *21* (9), 4375–4385. https://doi.org/10.1021/ACS.JCTC.5C00340
(39) Tribello, G. A.; Bonomi, M.; Branduardi, D.; Camilloni, C.; Bussi, G. PLUMED 2: New Feathers for an Old Bird. *Comput. Phys. Commun.* **2014**, *185* (2), 604–613. https://doi.org/10.1016/J.CPC.2013.09.018
(40) Brunken, C.; Peltre, O.; Chomet, H.; Walewski, L.; McAuliffe, M.; Heyraud, V.; Attias, S.; Maarand, M.; Khanfir, Y.; Toledo, E.; Falcioni, F.; Bluntzer, M.; Acosta-Gutiérrez, S.; Tilly, J. Machine Learning Interatomic Potentials: Library for Efficient Training, Model Development and Simulation of Molecular Systems. **2025**.
(41) Wehrhan, L.; Walewski, L.; Bluntzer, M.; Chomet, H.; Tilly, J.; Brunken, C.; Acosta-Gutiérrez, S. MLIPAudit: A Benchmarking Tool for Machine Learned Interatomic Potentials. **2025**, 2025–2036.
(42) Schäfer, M. R.; Segreto, N.; Zills, F.; Holm, C.; Kästner, J. Apax: A Flexible and Performant Framework for the Development of Machine-Learned Interatomic Potentials. *J. Chem. Inf. Model.* **2025**, *65* (15), 8066–8078. https://doi.org/10.1021/ACS.JCIM.5C01221
(43) Liu, H.; Zhu, L.; Zhang, X.; Yang, N.; Zhou, J.; Sun, Z. AI2Pot: A Scalable and Unified Framework for Machine-Learning Interatomic Potential Development and Large-Scale Molecular Dynamic Simulations. **2026**.
(44) Sharma, M.; Punnathanam, S.; Rajan, A. G. MLIP Studio: An Open Platform for Interactive Benchmarking and Atomistic Simulations Using Machine Learning Interatomic Potentials. **2026**.
(45) Paszke, A.; Gross, S.; Massa, F.; Lerer, A.; Bradbury Google, J.; Chanan, G.; Killeen, T.; Lin, Z.; Gimelshein, N.; Antiga, L.; Desmaison, A.; Xamla, A. K.; Yang, E.; Devito, Z.; Raison Nabla, M.; Tejani, A.; Chilamkurthy, S.; Ai, Q.; Steiner, B.; Facebook, L. F.; Facebook, J. B.; Chintala, S. PyTorch: An Imperative Style, High-Performance Deep Learning Library. *Adv. Neural Inf. Process. Syst.* **2019**, *32*.
(46) Bradbury, J.; Frostig, R.; Hawkins, P.; Johnson, M. J.; Katariya, Y.; Leary, C.; Maclaurin, D.; Necula, G.; Paszke, A.; VanderPlas, J.; Wanderman-Milne, S.; Zhang, Q. JAX: Composable Transformations of Python+NumPy Programs. 2018.
(47) Zills, F.; Schäfer, M. R.; Segreto, N.; Kästner, J.; Holm, C.; Tovey, S. Collaboration on Machine-Learned Potentials with IPSuite: A Modular Framework for Learning-on-the-Fly. *J. Phys. Chem. B* **2024**, *128* (15), 3662–3676. https://doi.org/10.1021/ACS.JPCB.3C07187
(48) Vandenhaute, S.; Cools-Ceuppens, M.; DeKeyser, S.; Verstraelen, T.; Van Speybroeck, V. Machine Learning Potentials for Metal-Organic Frameworks Using an Incremental Learning Approach. *npj Computational Materials* **2023**, *9* (1), 19-. https://doi.org/10.1038/s41524-023-00969-x
(49) Gelžinytė, E.; Wengert, S.; Stenczel, T. K.; Heenen, H. H.; Reuter, K.; Csányi, G.; Bernstein, N. Wfl Python Toolkit for Creating Machine Learning Interatomic Potentials and Related Atomistic Simulation Workflows. *Journal of Chemical Physics* **2023**, *159* (12). https://doi.org/10.1063/5.0156845/2913657
(50) Liu, Y.; Morrow, J. D.; Ertural, C.; Fragapane, N. L.; Gardner, J. L. A.; Naik, A. A.; Zhou, Y.; George, J.; Deringer, V. L. An Automated Framework for Exploring and Learning Potential-Energy Surfaces. *Nature Communications* **2025**, *16* (1), 7666-. https://doi.org/10.1038/s41467-025-62510-6
(51) Kühne, T. D.; Iannuzzi, M.; Del Ben, M.; Rybkin, V. V.; Seewald, P.; Stein, F.; Laino, T.; Khaliullin, R. Z.; Schütt, O.; Schiffmann, F.; Golze, D.; Wilhelm, J.; Chulkov, S.; Bani-Hashemian, M. H.; Weber, V.; Borštnik, U.; Taillefumier, M.; Jakobovits, A. S.; Lazzaro, A.; Pabst, H.; Müller, T.; Schade, R.; Guidon, M.; Andermatt, S.; Holmberg, N.; Schenter, G. K.; Hehn, A.; Bussy, A.; Belleflamme, F.; Tabacchi, G.; Glöß, A.; Lass,

M.; Bethune, I.; Mundy, C. J.; Plessl, C.; Watkins, M.; VandeVondele, J.; Krack, M.; Hutter, J. CP2K: An Electronic Structure and Molecular Dynamics Software Package - Quickstep: Efficient and Accurate Electronic Structure Calculations. *Journal of Chemical Physics* **2020**, *152* (19). https://doi.org/10.1063/5.0007045/199081

(52) Himanen, L.; Jäger, M. O. J.; Morooka, E. V.; Federici Canova, F.; Ranawat, Y. S.; Gao, D. Z.; Rinke, P.; Foster, A. S. DScribe: Library of Descriptors for Machine Learning in Materials Science. *Comput. Phys. Commun.* **2020**, *247*, 106949. https://doi.org/10.1016/J.CPC.2019.106949

(53) Liu, F. T.; Ting, K. M.; Zhou, Z. H. Isolation Forest. *Proceedings - IEEE International Conference on Data Mining, ICDM* **2008**, 413–422. https://doi.org/10.1109/ICDM.2008.17

(54) Lloyd, S. P. Least Squares Quantization in PCM. *IEEE Trans. Inf. Theory* **1982**, *28* (2), 129–137. https://doi.org/10.1109/TIT.1982.1056489

(55) Busk, J.; Schmidt, M. N.; Winther, O.; Vegge, T.; Jørgensen, P. B. Graph Neural Network Interatomic Potential Ensembles with Calibrated Aleatoric and Epistemic Uncertainty on Energy and Forces. *Physical Chemistry Chemical Physics* **2023**, *25* (37), 25828–25837. https://doi.org/10.1039/D3CP02143B

(56) Ho, C. H.; Ortner, C.; Wang, Y. S. Flexible Uncertainty Calibration for Machine-Learned Interatomic Potentials. *npj Computational Materials* **2026**, *12* (1), 225-. https://doi.org/10.1038/s41524-026-02080-3

(57) Kulichenko, M.; Barros, K.; Lubbers, N.; Li, Y. W.; Messerly, R.; Tretiak, S.; Smith, J. S.; Nebgen, B. Uncertainty-Driven Dynamics for Active Learning of Interatomic Potentials. *Nature Computational Science* **2023**, *3* (3), 230–239. https://doi.org/10.1038/s43588-023-00406-5

(58) Stark, W. G.; van der Oord, C.; Batatia, I.; Zhang, Y.; Jiang, B.; Csányi, G.; Maurer, R. J. Benchmarking of Machine Learning Interatomic Potentials for Reactive Hydrogen Dynamics at Metal Surfaces. *Mach. Learn. Sci. Technol.* **2024**, *5* (3), 030501. https://doi.org/10.1088/2632-2153/ad5f11

(59) Christensen, A. S.; von Lilienfeld, O. A. On the Role of Gradients for Machine Learning of Molecular Energies and Forces. *Mach. Learn. Sci. Technol.* **2020**, *1* (4), 045018. https://doi.org/10.1088/2632-2153/abba6f

(60) Kovács, D. P.; Oord, C. Van Der; Kucera, J.; Allen, A. E. A.; Cole, D. J.; Ortner, C.; Csányi, G. Linear Atomic Cluster Expansion Force Fields for Organic Molecules: Beyond RMSE. *J. Chem. Theory Comput.* **2021**, *17* (12), 7696–7711. https://doi.org/10.1021/ACS.JCTC.1C00647

(61) Abideen, Z. U.; Fernández-Pendás, M.; Rodriguez Quintero, C.; Usman, M.; Infante, I. Orchestr.AI v0.2.0: Unified Platform for Machine-Learning Force Fields. https://doi.org/10.5281/ZENODO.22894956